\documentclass{moriond}

\def\Journal#1#2#3#4{{#1} {\bf #2}, #3 (#4)}

\def\be{\begin{equation}}
\def\ee{\end{equation}}
\def\bea{\begin{eqnarray}}
\def\eea{\end{eqnarray}}

\newcommand{\Photo}{\includegraphics[height=35mm]{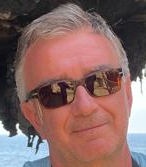}}
\usepackage{amsmath}

\begin{document}
\vspace*{4cm}

\title{MEASUREMENT OF THE PROTON MAXIMUM ACCELERATION ENERGY IN GALACTIC COSMIC RAYS}

\author{G. DI SCIASCIO }

\address{INFN - Roma Tor Vergata, Viale della Ricerca Scientifica 1, Roma, Italy}

\maketitle\abstracts{Cosmic rays represent one of the most important energy transformation processes of the universe.
They bring information about the surrounding universe, our galaxy, and very probably also the extragalactic space, at least at the highest observed energies. 
More than one century after their discovery, we have no definitive models yet about the origin, acceleration and propagation processes of the radiation.
The main reason is that there are still significant discrepancies among the results obtained by different experiments, probably due to some still unknown systematic uncertainties affecting the measurements.
In this paper, we will focus on the detection of galactic cosmic rays in the 10$^{15}$ eV energy range, where the so-called \emph{`knee'} in the all-particle energy spectrum is observed. The measurement of the (p+He) energy spectrum presented and discussed.
}

\section{Introduction}

The origin of Cosmic Rays (CRs) is still an open problem.
The observation of the diffuse gamma-ray emission from the galactic disk is the strongest demonstration that the bulk of CRs are Galactic, together with the observation that the gamma flux from LMC is weaker than expected according to the mass of the target \cite{ackerman2016}.
In fact, in 1962 Hayakawa suggested that CR protons meet the InsterStellar Medium (ISM) producing photons through the decay of neutral pions after a pp or a p$\gamma$ interaction, with the gamma flux roughly scaling with the amount of crossed material \cite{hayak1962}. 
In addition, magnetic fields at work in the Galaxy can confine CRs with energy E if the Larmor radius of these particles, $r_L(E)$, is smaller than the dimension of the galactic halo $H$,  $r_L(E)\leq H$. If the magnetic field in the ISM is $\sim\mu$G and $H\sim$kpc we obtain that CRs protons with energy $E\leq$10$^{16-17}$ eV could be confined in the Galaxy. 

These considerations are the basis of the CR standard model:  the bulk of CRs up to about 10$^{17}$ eV are Galactic, produced and accelerated by the shock waves of SuperNova Remnants (SNR) expanding shells \cite{drury12}, 
The particles diffuse over very long time (a few Myears) inside the Galaxy under the effects of the magnetic fields which scramble their arrival direction distribution. 
A transition to an extragalactic component is expected somewhere between 10$^{17}$--10$^{19}$ eV.

\begin{figure}[t]
  \begin{center}
  {\includegraphics[width=0.9\linewidth]{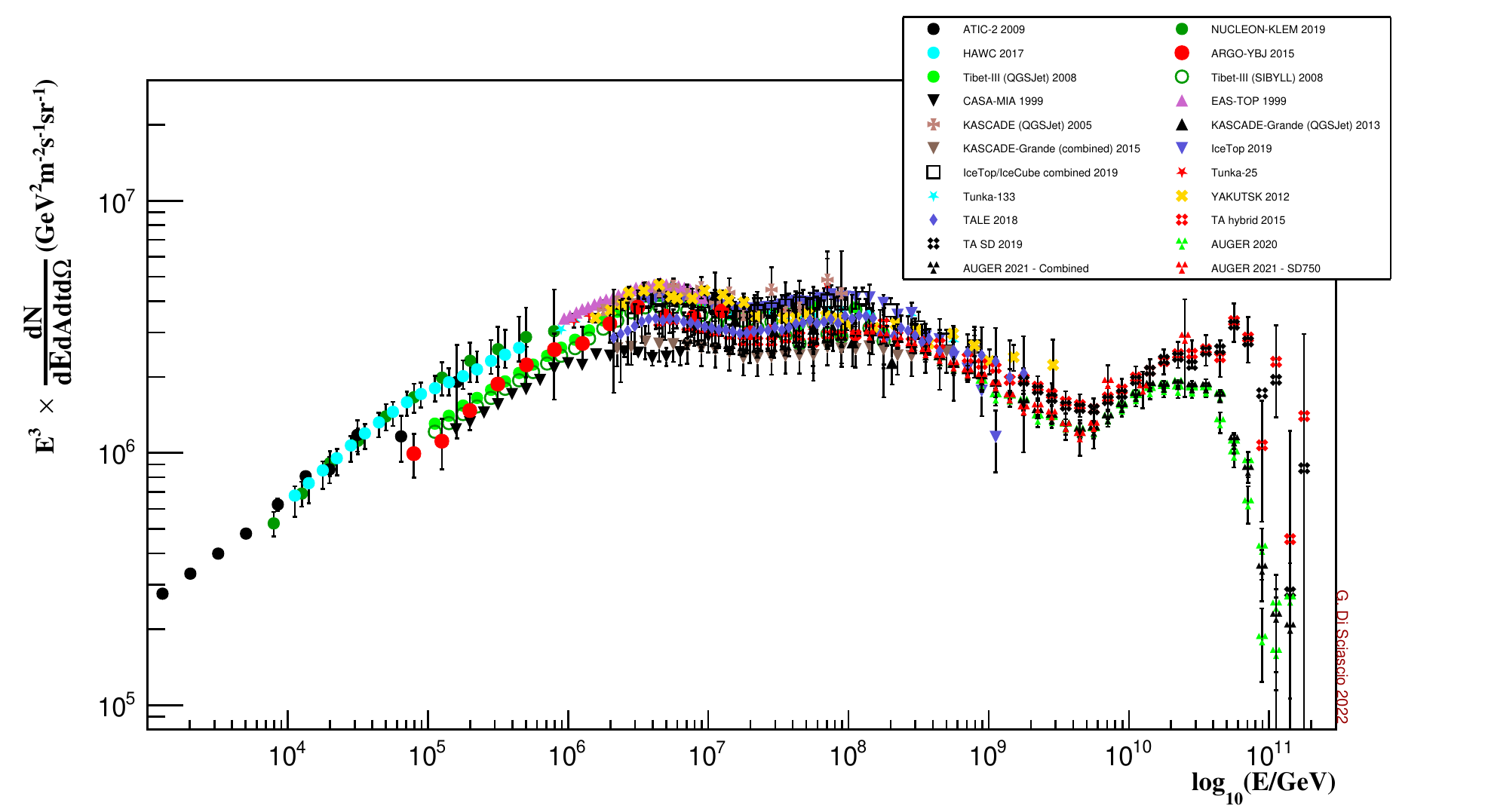}}
  \vspace{-0.5pc}
\caption{All-particle energy spectrum of primary cosmic rays measured by different experiments.}
\label{fig:allpart-enespt}
  \end{center}
\end{figure}

The primary CR all-particle energy spectrum (namely the number of nuclei as a function of total energy) exceeds 10$^{20}$ eV and is shown in Fig. \ref{fig:allpart-enespt}.
On general grounds, the spectrum can be described by different simple power laws ($\sim K\cdot E^{-\gamma}$) in adjacent energy intervals separated by some features consisting in more or less rapid change of the spectral index $\gamma$. These features can be described as a series of hardening ("knee-like") or softening ("ankle-like") of the energy spectrum. 
Despite the differences in flux, emphasized by multiplying the differential spectrum by E$^3$, all the measurements of the all-particle energy spectrum are in fair agreement when taking into account the statistical, systematic and energy scale uncertainties. They all show a few basic characteristics:
%
\begin{enumerate}
\item[(a)] a power-law behaviour $\sim$ E$^{-2.7}$ until the so-called \emph{``knee''}, a small downwards bend around few PeV; 
\item[(b)] a power-law behaviour $\sim$ E$^{-3.1}$ beyond the knee, with a slight dip near 10$^{17}$ eV, sometimes referred to as the \emph{``second knee''}; 
\item[(c)] a transition back to a power-law $\sim$ E$^{-2.7}$ (the so-called \emph{``ankle''}) around $4-5\cdot10^{18}$~eV; 
\item[(d)] a cutoff probably due to extra-galactic CR interactions with the Cosmic Microwave Background (CMB) around 10$^{20}$ eV (the Greisen-Zatsepin-Kuzmin effect).
\end{enumerate}
%
%
All the observed features are believed to carry fundamental information that sheds light on the key questions of the origin, acceleration and propagation of CRs. However, from the all-particle results alone, it is not possible to understand the origin of different features. All models concerning sources, acceleration and propagation of the primary flux, differ considerably for what concerns expected elemental composition as a function of the energy. A measurement of the chemical composition is therefore crucial to disentangle between different hypotheses.

Around 10$^{15}$ eV the flux is so low (about 1 particle/m$^2$/year) that the only chance to have a statistically significant detection is to built earth-based detectors of large area, operating for long times. In that case, the atmosphere is considered as a target, and we study the primary properties in an \emph{`indirect'} way, through the measurement of secondary particles produced in the interaction of the primary particle with the nuclei of the atmosphere, the so-called \emph{`Extensive Air Shower'} (EAS).

Approaching the hundred TeV energy region, even in space-borne experiments, the energy assignment is indirect since it is generally based on the energy deposition of particles produced in the interaction of primaries in the detector itself. The reconstruction of the total energy is then obtained by comparison with some model prediction, and therefore, at least in that region, the boundary line between `direct' and `indirect' experiments is more uncertain.

The great variety of detection levels, layouts, observables, and reconstruction procedures to infer the elemental composition is at the origin, in part, of the conflicting results reported by different ground-based experiments above 100 TeV \cite{disciascio2022}.
Arrays focused on the investigation of the knee region operated so far are also characterized by a limited size of the instrumented area. They collected limited statistics above 10$^{16}$ eV, and were, therefore, unable to give a conclusive answer to the origin of the knee. In fact, the poor sensitivity to elemental composition, due to the small statistics, prevents  discrimination against different mass groups, and only general trends can be investigated in terms of the evolution of $\left<\ln A\right>$ or of \emph{``light''} and \emph{``heavy''} components with energy.
%
\begin{figure}[t]
  \begin{center}
  {\includegraphics[width=0.9\linewidth]{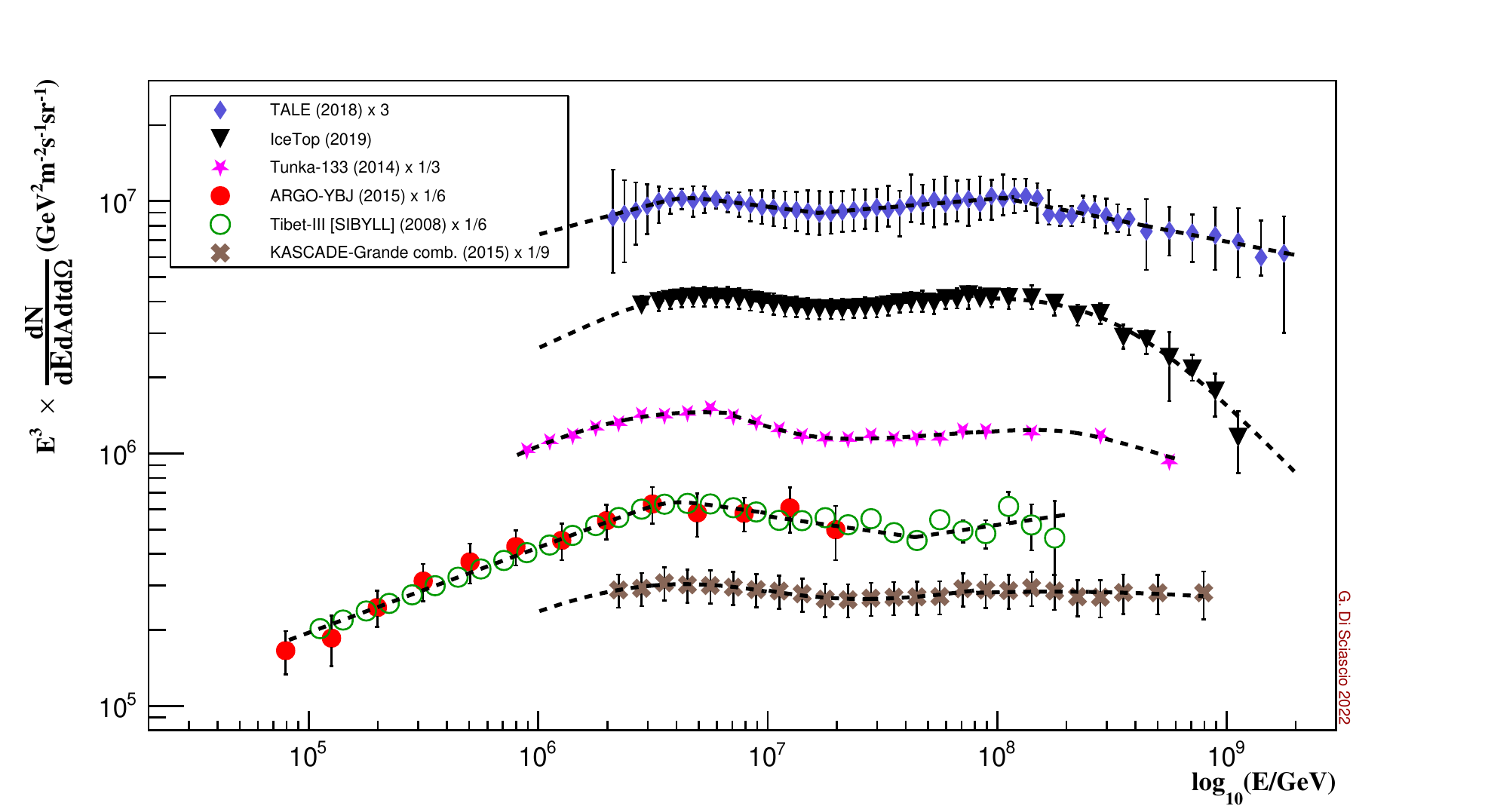}}
  \vspace{-0.5pc}
\caption{All-particle energy spectra of primary cosmic rays measured by TALE,
IceTop, 
Tunka-133, 
ARGO-YBJ, 
Tibet AS$\gamma$, 
KASCADE-Grande 
experiments. 
The total errors are plotted. The lines are fits to the different spectra with the formula (\ref{eq:spect-parametriz}). Best fits parameters are reported in Table \ref{tab:spectr-fitparam}. }
\label{fig:allp-fit}
  \end{center}
  \end{figure}
%
\section{The "Knee" in the all-particle energy spectrum}

The main structure is the \emph{``Knee''} observed for the first time by R.W. Williams in 1948 in the experiment which first located individual shower cores from symmetry of the fired detectors \cite{williams1948,linsley1983}.
We do not know the origin of the knee but, as first suggested in 1959 by Kulikov and Khristiansen \cite{khristiansen1959}, this feature could be connected to the maximum energy at which Galactic CRs are accelerated and, consequently, to the end of the Galactic CR flux. 

In 1961 Peters \cite{peters1961} proposed a rigidity-based scheme to explain the mechanism of particle acceleration. According to this model protons will cutoff first, followed by other nuclei according to the relation
\begin{equation}
E_\text{max}(Z) = Z\times E_\text{max} (Z=1).
\end{equation}
If the dominant primary mass of the knee is light (protons and helium), then, according to this scheme, the Galactic CR spectrum is expected to end with iron at about 10$^{17}$ eV, where a second knee is observed.
The different experiments observe a knee at about 3-4 PeV, a small softening at about 10-15 PeV and a second knee around 200 -- 300 PeV (see Fig. \ref{fig:allpart-enespt}).

To quantify the different features in this energy range we described the energy spectrum measured by ARGO-YBJ \cite{disciascio-rev}, Tibet AS$\gamma$ \cite{amenomori2008}, Tunka-133 \cite{tunka133-allp}, KASCADE-Grande \cite{schoo2015}, TALE \cite{tale2018-allp} and IceTop \cite{aartsen2019} experiments with a widely used form \cite{terantonyan2000,schatz2001,horandel2003,lipari2018,lipari2020}
\begin{equation}
 \phi(E) = K_0 \;
 \left ( \frac{E}{E_0} \right )^{-\gamma_1} \;
 \left [1 + \left (\frac{E}{E_b} \right )^{\frac{1}{w}} \right ]^{-(\gamma_2 - \gamma_1) \, w}
\label{eq:spect-parametriz}
\end{equation}
The absolute flux $K_0$ and the spectral index $\gamma_1$ quantify the power law. $E_0$ is a reference energy. The flux above the cut-off energy $E_b$ is modeled by a second and steeper power law. The parameters $\gamma_2$, the slope beyond the knee, and $w>0$, the smoothness of the transition from the first to the second power law, characterize the change in the spectrum at the cut-off energy. A value $w=0$ corresponds to a steep transition that soften with increasing values \cite{lipari2018}.
%
\begin{table}
  \caption{\footnotesize Fits to the all--particle CR spectra in the energy range $8\cdot 10^4$ to $2\cdot 10^9$~GeV.}
    \label{tab:spectr-fitparam}
  \vspace{0.2 cm}
{\footnotesize  (a) Parameters for the first Knee.  }
  
\begin{center}
\begin{tabular}{ | l || c |  c |  c | c |}
\hline
Experiment   &  $E_{b1}$  (PeV)  &  $\gamma_1 $   &  $\gamma_2$  &   $w_1$  \\
\hline
TALE          & 4.26 $\pm$ 1.65     & 2.76 $\pm$ 0.18  & 3.11 $\pm$ 0.07 & 0.07 $\pm$ 0.18 \\
IceTop        &  3.30 $\pm$ 1.23   &  2.48 $\pm$ 0.08  & 3.12 $\pm$ 0.12 & 0.30 $\pm$ 0.46 \\
Tunka--133   & 4.18 $\pm$ 0.83 & 2.76 $\pm$ 0.09   & 3.20 $\pm$ 0.04 & 0.15 $\pm$ 0.16 \\
ARGO--YBJ/Tibet AS$\gamma$  & 3.72 $\pm$ 0.03 & 2.66 $\pm$ 0.01 & 3.13 $\pm$ 0.01 & 0.11 $\pm$ 0.01 \\
Kascade--Grande    & 2.10 $\pm$ 0.87 & 2.47 $\pm$ 0.04 & 3.16 $\pm$ 0.14 & 0.60 $\pm$ 0.51 \\
\hline
\end{tabular}
\end{center}

  \vspace{0.2 cm}
{\footnotesize  (b) Parameters for the ankle feature. }

  \begin{center}
\begin{tabular}{ | l || c |  c |  c | c |}                                                
\hline
Experiment   &  $E_{b2}$  (PeV)  &  $\gamma_2 $   &  $\gamma_3$  &   $w_2$   \\   
\hline
TALE          & 16.61 $\pm$ 8.36     & 3.11 $\pm$ 0.05 & 2.93 $\pm$ 0.05 & 0.07 $\pm$ 0.05 \\
IceTop        &  18.66 $\pm$ 6.65    & 3.12 $\pm$ 0.12 & 2.92 $\pm$ 0.05 & 0.05 $\pm$ 0.05 \\
Tunka--133    & 18.70 $\pm$ 3.88 & 3.20 $\pm$ 0.04 & 2.96 $\pm$ 0.05 & 0.17 $\pm$ 0.45 \\
ARGO--YBJ/Tibet AS$\gamma$ & 43.8 $\pm$ 4.81 & 3.13 $\pm$ 0.01 & 2.86 $\pm$ 0.05 & 0.01 $\pm$ 0.01 \\
Kascade--Grande    & 18.01 $\pm$ 17.4 & 3.16 $\pm$ 0.14 & 2.83 $\pm$ 0.45 & 0.66 $\pm$ 1.74 \\
\hline
\end{tabular}
\end{center}

  \vspace{0.2 cm}
{\footnotesize  (c) Parameters for the second Knee. }

  \begin{center}
\begin{tabular}{ | l || c |  c |  c | c |}                                                
\hline
Experiment   &  $E_{b3}$  (PeV)  &  $\gamma_3 $   &  $\gamma_4$  &   $w_3$   \\   
\hline
TALE          & 104.5 $\pm$ 40.0     & 2.93 $\pm$ 0.05 & 3.18 $\pm$ 0.06 & 0.02 $\pm$ 0.02 \\
IceTop        &  168.4 $\pm$ 17.4  & 2.92 $\pm$ 0.05 & 3.50 $\pm$ 0.40 & 0.25 $\pm$ 0.16 \\
Tunka--133    & 238.2 $\pm$ 56.8  & 2.96 $\pm$ 0.05 & 3.34 $\pm$ 0.19 & 0.05 $\pm$ 0.50 \\
Kascade--Grande    & 274.5 $\pm$ 122 & 2.83 $\pm$ 0.45 & 3.20 $\pm$ 0.13 & 2.47 $\pm$ 0.97 \\
\hline
\end{tabular}
\end{center}
\end{table}
%

In Fig. \ref{fig:allp-fit} some selected measurements of the all--particle energy spectrum in the
energy region from $8\cdot 10^4$ to $2\cdot 10^9$~GeV are shown.
The data come from ARGO-YBJ \cite{disciascio-rev},Tibet AS$\gamma$ (Sibyll) \cite{amenomori2008}, Kascade-Grande \cite{schoo2015}, IceTop \cite{aartsen2019}, Tunka-133 \cite{tunka133-allp}, TALE \cite{tale2018-allp} experiments.
As it can be seen, ARGO-YBJ and Tibet AS$\gamma$ are the only shower arrays that traced the knee in detail, starting from more than an energy decade below. Instead the other experiments have an energy threshold too close to the knee.

Different spectra agree in showing a knee at a few PeV, an ankle right after and a second knee at about 200--300 PeV.
But the different experiments also show important differences related to large systematic errors.
By assuming the existence of these structures we described the spectra with the formula (\ref{eq:spect-parametriz}) summarizing the best fit parameters in Table~\ref{tab:spectr-fitparam}. 
The spectrum is described as four segments with constant spectral index, $\gamma_1$, $\gamma_2$ and $\gamma_3$, separated by three spectral features (a knee, an ankle and another knee) with break energies $E_{b1}$, $E_{b2}$ and $E_{b3}$ and widths $w_1$,  $w_2$ and $w_3$.
We used the total error, combining quadratically statistically and systematic errors.

As expected the most accurate determination of the first knee comes from the ARGO-YBJ/Tibet AS$\gamma$ global fit with $E_{b1}=3.72\pm 0.03$ PeV. The spectral index before the knee is $\gamma_1=2.66\pm 0.01$ and after $\gamma_2=3.13\pm 0.01$. The small ankle feature results at about 18 PeV with an index softening at about 3. In this case data from Tibet AS$\gamma$ are not statistically significant for a precise determination. The determination of the second knee is spread in the range 100--300 PeV with a hardening of the spectrum at about 3.3 .

\section{Measurement of the "Knee" in the (p+He) energy spectrum}

Several experimental results associate the all-particle knee with the bending of the light component (p+He), and are compatible with a rigidity-dependent cut-off \cite{aglietta2004,kascade2005,kascade2,kascade3,garyaka2007,tanaka2012}. However, in some experiments the flux of the different components vary significantly depending on the hadronic interaction model used to interpret the data \cite{kascade2005,kascade2,kascade3}.
On the contrary, other results (in particular those obtained by arrays located at high altitudes) seem to indicate that  the knee of the all-particle energy spectrum is due to heavier nuclei and that the light component cuts off below \mbox{1 PeV  \cite{disciascio-rev,aglietta2004,argo-hybrid2,amenomori2006,casamia,basje-mas}}. 

\begin{figure}[h]
\begin{center}
\includegraphics[width=0.9\textwidth]{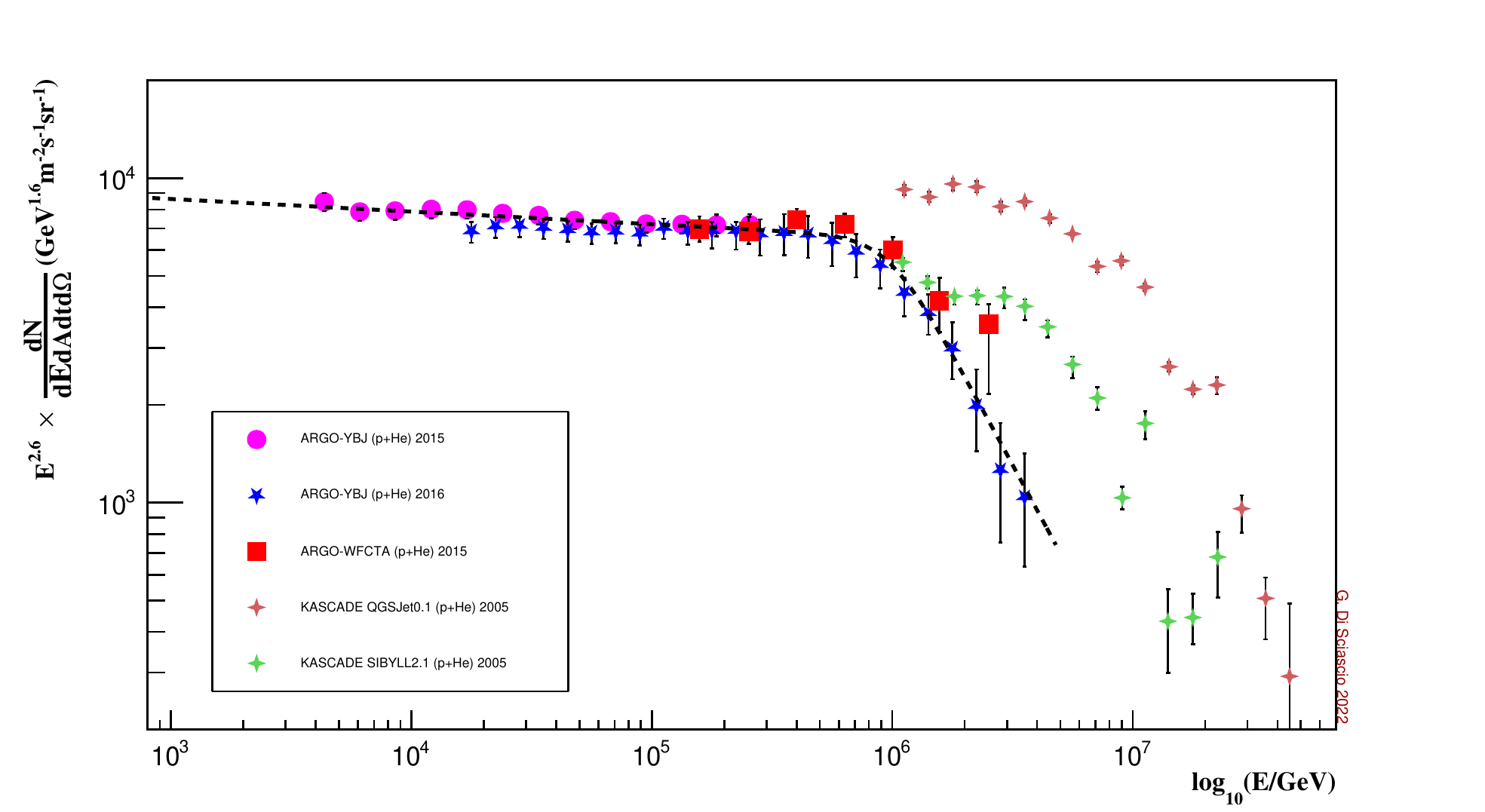}
\caption{Energy spectra of the light (p+He) component as measured by the ARGO-YBJ 
experiment with different techniques and analyses, compared with results obtained by KASCADE with two different interaction models
The dashed line shows a global fit to all ARGO-YBJ results.}
 \label{fig:pheknee}
    \end{center}
    \end{figure}

In Figure~\ref{fig:pheknee} the energy spectra of the light component as measured by  the ARGO-YBJ \cite{disciascio-rev} and KASCADE \cite{kascade2005,kascade2,kascade3} experiments are shown. 
This plot exemplifies the conflicting results in the knee region.
 ARGO-YBJ shows evidence of a (p+He) cutoff below the PeV, KASCADE on the other hand reports a light knee at few PeV, consistent with the position of the all-particle knee.

ARGO-YBJ, located at 4300 m a.s.l., did not exploit a measurement of the muon component to determine the elemental composition of the primary CR flux. 
On the contrary, KASCADE, located at sea level, used the N$_e$/N$_{\mu}$ correlation to select different primary masses with two different hadronic interaction models \cite{kascade2005,kascade2,kascade3}. 

ARGO-YBJ is the only experiment that traced the (p+He) component across the knee starting from an energy so low ($\approx$TeV) to overlap with direct measurements thus cross-calibrating the fluxes on a wide energy range (5--250 TeV).
The cross-calibration of fluxes in this energy range, where the boundary line between `direct' and `indirect' measurements is uncertain, is very important.
The low energy threshold allowed also a calibration of the absolute energy scale at a level of 10\% exploiting the \emph{Moon Shadow} technique in the 1--30 TeV/Z range \cite{bartoli2011}.
The energy threshold of KASCADE is, on the contrary, about 1 PeV.
%
%

The ARGO-YBJ experiment measured the CR energy spectra exploiting completely different and independent methods \cite{disciascio-rev,bartoli2011,bartoli2012,bartoli15}:
\begin{itemize}
\item \emph{`Digital-Bayes' analysis}, based on the strip multiplicity, that is, the picture of the EAS provided by the RPC strip/pad system, in the few TeV--300 TeV energy range. The selection of light elements is based on the characteristics of the charged particle lateral distribution \cite{bartoli2011,bartoli2012,bartoli15,bartoli17}.
\item \emph{`Analog-Bayes' analysis}, based on the RPC charge readout \cite{argo-bigpad}, covers the 30 TeV--10~PeV range. The energy is reconstructed, as in the previous analysis, by using a bayesian approach.
\item \emph{`Hybrid measurement'}, carried out exploiting also a wide field of view Cherenkov telescope, a prototype of the LHAASO telescopes, in the 100 TeV--3 PeV region. The selection of (p+He)-originated showers is based on two observables, the shape of the Cherenkov image and the particle density in the core region measured by the ARGO-YBJ central carpet. The energy is reconstructed by the telescope with a resolution better than 20\% \cite{argo-hybrid1,argo-hybrid2}.
\end{itemize}
As it can be seen from Fig. \ref{fig:pheknee}, all ARGO-YBJ results are in good agreement.
The contamination of nuclei heavier than helium is estimated smaller than 15\% at 1 PeV in all analyses.
%
\begin{figure}[h]
\begin{center}
\includegraphics[width=0.9\textwidth]{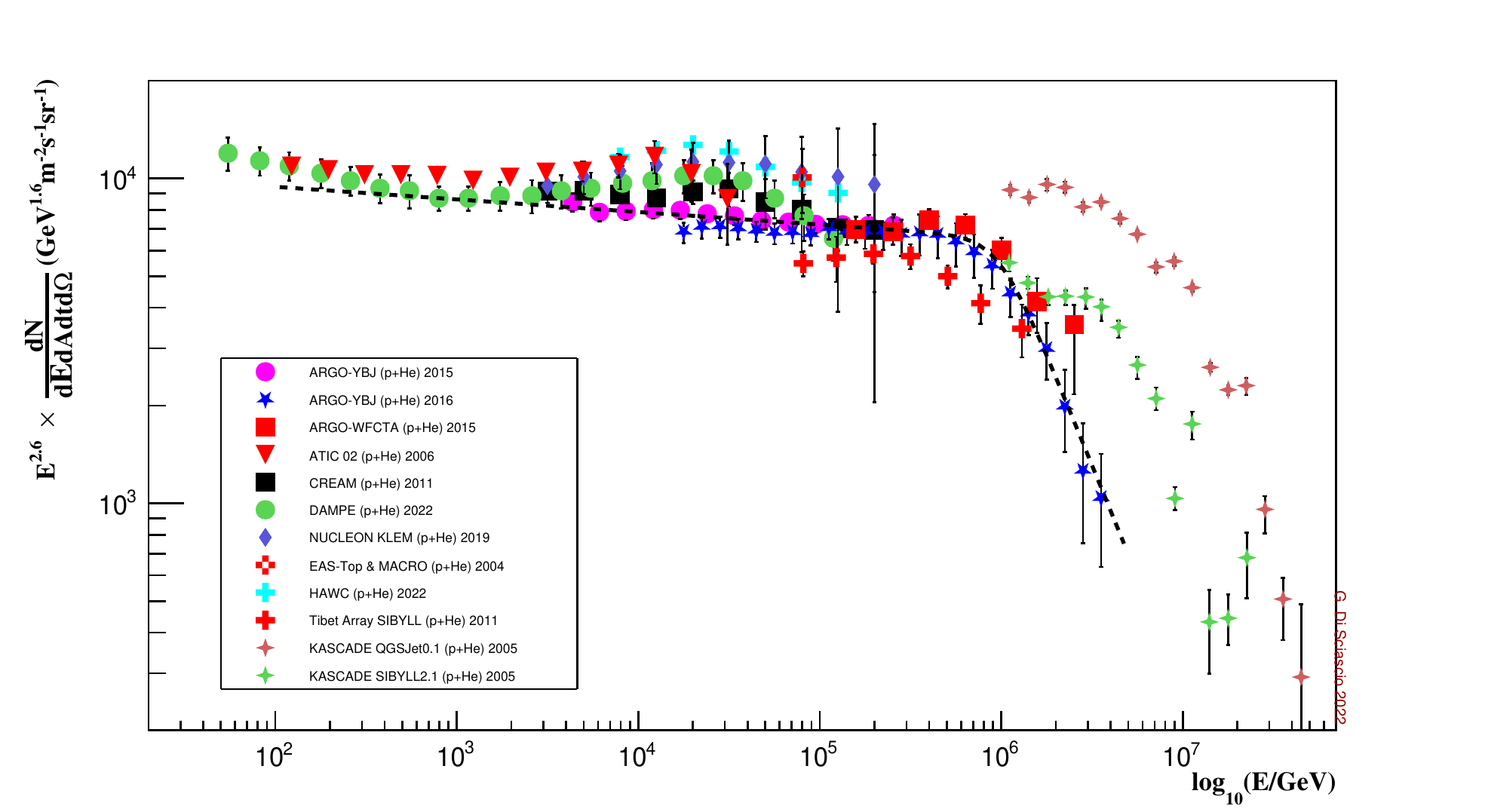}
\caption{Energy spectra of the light (p+He) component as measured by ARGO-YBJ and KASCADE compared with data 
from HAWC, EAS-TOP and Tibet AS$\gamma$ air shower arrays, and with measurements from direct detectors ATIC, CREAM, NUCLEON and DAMPE. } 
 \label{fig:light-allexp}
    \end{center}
    \end{figure}   
%
%
\begin{figure}[h]
\begin{center}
\includegraphics[width=0.9\textwidth]{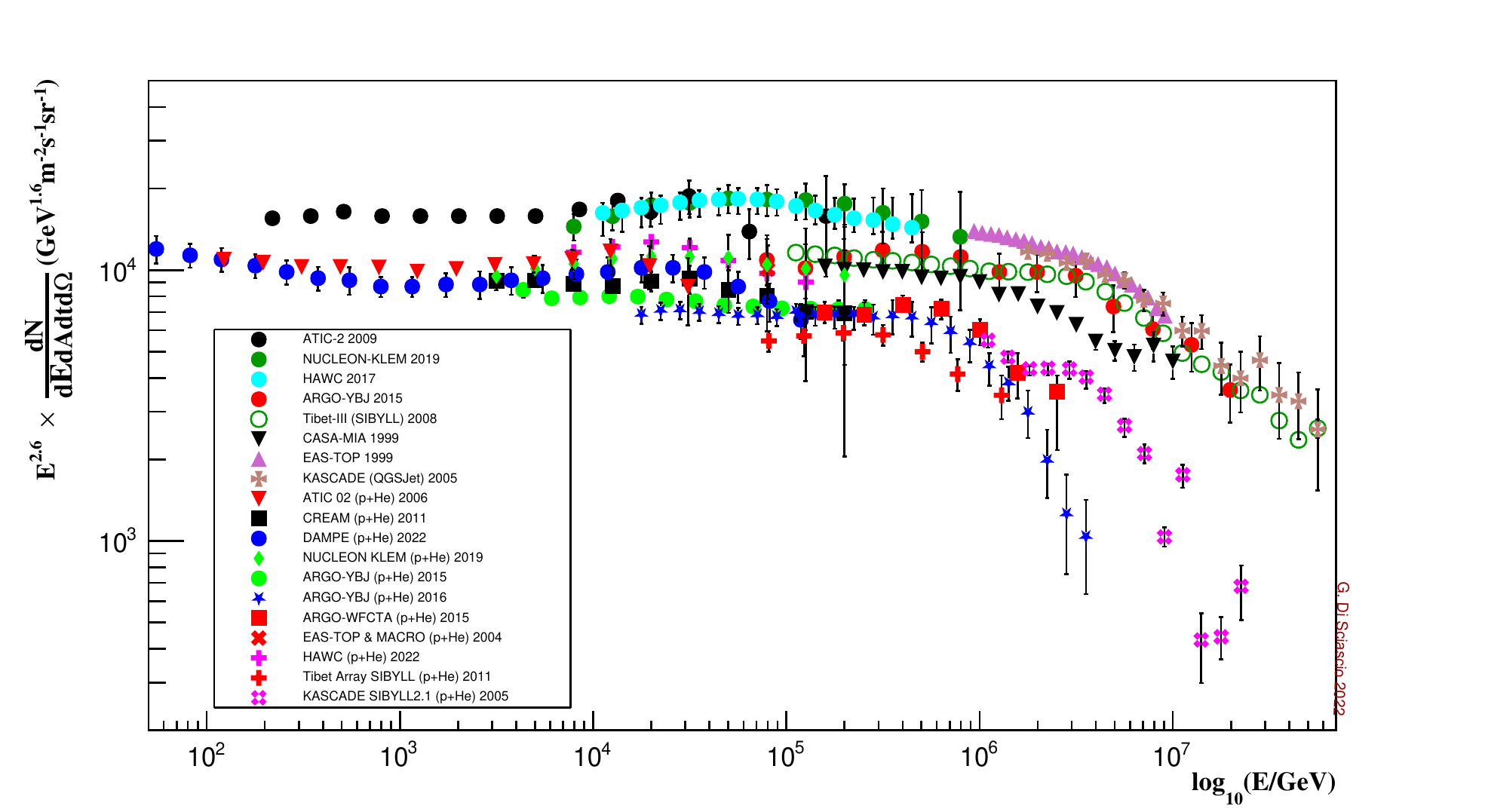}
\caption{All-particle and light (p+He) energy spectra measured by a selection of detectors. } 
 \label{fig:knee-allp-light}
    \end{center}
    \end{figure}   
%

A global fit to the ARGO-YBJ measurements with the formula (\ref{eq:spect-parametriz}) gives the following results: $E_b = 914\pm 259$ TeV, $\gamma_1 = 2.64\pm 0.01$, $\gamma_2 = 3.98\pm 0.42$, $w = 0.20\pm 0.13$, showing the (p+He) energy spectrum can be described by a single power law in the range TeV -- 800 TeV.

In Fig. \ref{fig:light-allexp} the ARGO-YBJ and KASCADE data are compared with results obtained by other air shower arrays HAWC \cite{hawc-light}, EAS-TOP \cite{aglietta2004} and Tibet AS$\gamma$\cite{amenomori2008}, and with measurements from direct detectors ATIC \cite{atic}, CREAM \cite{cream2011}, NUCLEON \cite{nucleon2019} and DAMPE \cite{dampe}. 
Around 10 TeV ARGO-YBJ, DAMPE, CREAM, NUCLEON are in fair agreement within the systematic uncertainties.
Extrapolations of ARGO-YBJ results are in fair agreement with DAMPE down to about 100 GeV.
Close to 100 TeV all experiments are in agreement within the systematic uncertainties.
These results show that, when indirect measurements are capable of selecting almost pure beams, their findings are in fair agreement with direct ones and confirm that current simulation models provide a satisfactory description of the EAS development in the atmosphere.

Nevertheless, in the 10 -- 100 TeV energy range HAWC, ATIC, DAMPE and NUCLEON fluxes are higher than ARGO-YBJ showing evidence of a possible deviation from a single power law behaviour.
None of these experiments made measurements up to the knee of the all-particle energy spectrum to calibrate the fluxes and only NUCLEON, ATIC and HAWC measured the all-particle spectrum reporting a flux higher than ARGO-YBJ and Tibet AS$\gamma$. 

Above 100 TeV ARGO-YBJ and Tibet AS$\gamma$ are in good agreement showing both a cut off in the (p+He) component below 1 PeV.

\section{Conclusions and outlook}

The results obtained in the knee region, starting from about 10 TeV, are conflicting, in particular for what concern the elemental composition. Evidence for a deviation from a single power law in the 10--100 TeV range is reported by different experiments. The observation of a (p+He) cut-off below the knee quoted by ARGO-YBJ, Tibet AS$\gamma$, CASA-MIA, BASJE-MASS is at variance with results reported by other experiments such as  KASCADE, IceTop, Tunka.
In Fig. \ref{fig:knee-allp-light} a selection of all-particle and light component energy spectra in the knee region is shown. 
Some (p+He) fluxes are at level of the all-particle fluxes measured by other experiments.
The situation is still quite confusing and it will be interesting if HAWC will extend its measurements up to the all-particle knee to calibrate the fluxes.

New measurements able to trace at least the light and heavy components in the energy range $10^{12} - 5\cdot 10^{17}$ eV with high statistics are therefore needed.
The only experiment that meets these requirement is LHAASO, a new multi-component array located at high altitude (4410 m asl, 600 g/cm$^2$) in the Daochen site, Sichuan province, P.R. China. 
LHAASO is expected to measure the energy spectrum, the elemental composition and the anisotropy of CRs starting from the TeV range, thus overlapping direct measurements in a wide interval~\cite{lhaaso-wb}.
The experiment is constituted by a 1 km$^2$ dense array of plastic scintillators and muon detectors. At the center of the array a 300~$\times$~300 m$^2$ water Cherenkov facility will allow the detection of TeV showers. An array of 18 wide field of view Cherenkov telescopes will image the longitudinal profile of events. With a 40,000 m$^2$ area muon detector LHAASO will study the muonic component and the $N_e/N_{\mu}$ correlation with unprecedented sensitivity and resolution.

\end{document}